# Ultrafast Topological Transitions Driven by Permittivity Modulation in Non-Hermitian Multilayers

Giuseppina Simone

**Abstract.** Ultrafast permittivity modulation in epsilon-near-zero (ENZ) media provides a pathway for real-time control of non-Hermitian photonic topology. We model ultrafast topological dynamics in an ITO/SiO$_2$/Ag multilayer supporting hybrid epsilon-near-zero (ENZ)‐plasmon modes. Using a time-dependent Drude‐Lorentz permittivity for ITO and rigorous coupled-wave analysis, it is found that carrier relaxation with a 10-ps time constant redshifts the ENZ resonance from 490 to 525 nm and shifts the reflection-phase singularities across the angle‐wavelength plane. The motion of these singularities corresponds to the displacement of exceptional points and a discrete change in the cumulative winding number. The transient Floquet-like band structure exhibits avoided crossings and Dirac-type dispersions driven by ENZ modulation. The results directly link picosecond scale permittivity dynamics to quantized topological transitions in non-Hermitian photonic systems.

All-optical control of photonic states underpins ultrafast signal processing, quantum communication, and reconfigurable nanophotonics [1,2]. Conventional materials offer weak nonlinearities, demanding high switching energies and bulky devices [3,4]. ENZ media provide an alternative pathway by enabling extreme light–matter interactions, large refractive-index shifts, and strong nonlinearities on picosecond timescales [5,6]. Transparent conducting oxides such as indium tin oxide (ITO) combine CMOS compatibility, tunable carrier densities, and plasmonic behavior, establishing them as a leading ENZ platform [7]. Embedding ITO within resonant architectures further amplifies the nonlinear response, allowing ultrafast and reversible modulation [8]. Despite similar advances, the connection between picosecond-scale ENZ dynamics and non-Hermitian topological phenomena–such as exceptional points and phase singularities, remains largely unexplored [9]. Recent work has shown that prism-coupled ENZ films can achieve near-perfect absorption and broadband reflection switching using picosecond excitation [10-12], emphasizing the potential of ENZ media for ultrafast nonlinear photonics. However, standard reflectance spectra reveal only resonance energies, obscuring how eigenmodes connect and how singularities evolve [13,14]. In contrast, the cumulative reflection phase provides a continuous topological observable that reveals phase vortices with quantized winding numbers, robust invariants of the underlying non-Hermitian system [15]. Temporal modulation can displace the vortices and induce encounters with exceptional points, producing discrete jumps in winding number even as the material parameters vary smoothly. In this work, we investigate a planar multilayer that integrates an ENZ layer within a plasmonic–dielectric, metal configuration, supporting hybrid ENZ, plasmon modes with strong temporal tunability. Using a time-dependent

Drude–Lorentz model and rigorous coupled-wave analysis, we show that carrier relaxation in ITO displaces topological singularities and drives exceptional-point transitions, leading to quantized topological changes on picosecond timescales. The results establish a direct link between ultrafast material dynamics and discrete topological evolution, outlining a practical route toward reconfigurable non-Hermitian photonic devices and singularity-based optical control [16,17].

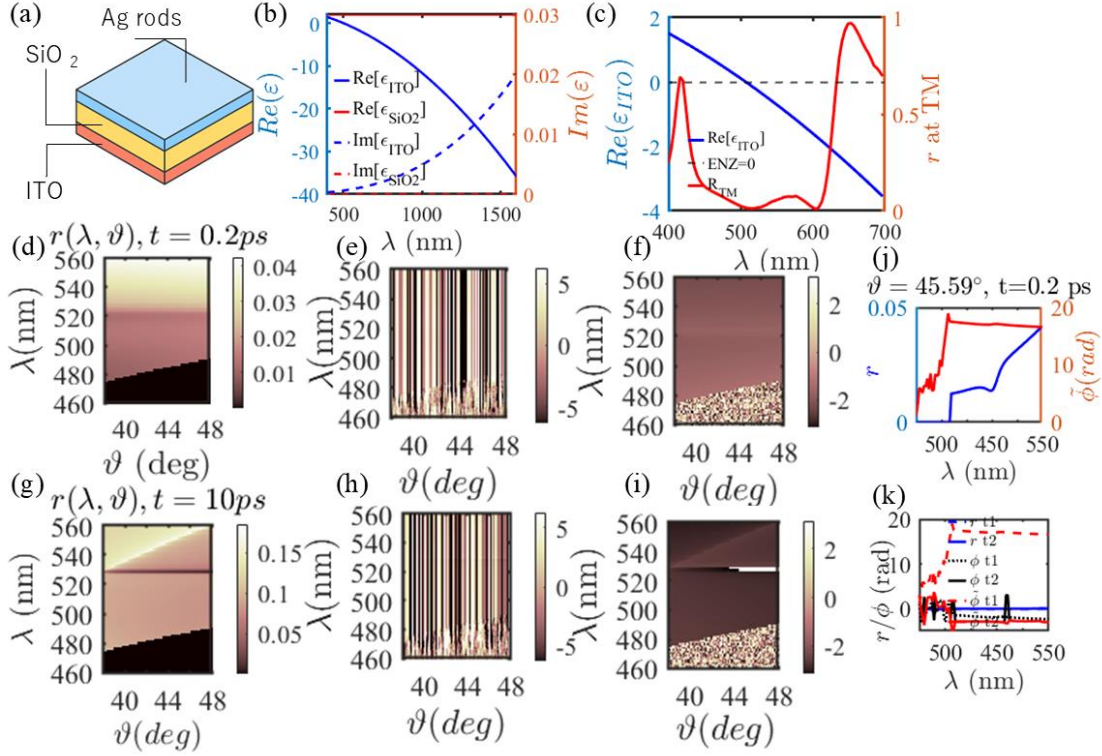

Figure 1. Hybrid ENZ–plasmonic time dependent resonance. (a) Schematic of multilayer (b–c) $\varepsilon_{ITO}$ and field coupling with ENZ @ 520 nm. (d,g) $\lambda$–$\vartheta$ reflectance maps, $t = 0.2$ and $10\ ps$. (e,h) $2\pi$-modulo phase of the complex reflectance $arg[r]\ (-\pi, \pi]$ (color bar $\pm 6.28$). (f,i) Cumulative unwrapped phase $\tilde{\phi}$. (j,k) Cross-sections @ $\vartheta = 45.6°$ comparing reflectance and phase at the two delays.

We model a planar ITO/SiO$_2$/Ag multilayer supporting hybrid epsilon-near-zero (ENZ)–plasmon modes (Figs. 1a). The Ag layer is composed of microrods [18-20], which introduce symmetry breaking and enable a dynamically tunable topological phase. Using rods instead of a continuous film enhances modal coupling and breaks in-plane symmetry. Carrier relaxation in the ITO layer is described by a time-dependent Drude–Lorentz permittivity, and the reflectance spectra are computed using rigorous coupled-wave analysis, Appendix A). From the complex reflection coefficient, both amplitude and phase, enabling identification of phase vortices and exceptional points, are extracted. Quantized winding numbers are obtained by evaluating the phase evolution around closed parameter loops, where discontinuous jumps indicate topological transitions. Exceptional points correspond to spectral degeneracies where eigenvalues coalesce, and the eigenvector matrix becomes ill-conditioned. Full derivations of the winding-number algorithm, exceptional-point tracking, and validation tests are provided in Appendix A. The architecture of the multilayer supports hybrid plasmonic resonances at the interfaces. Coupling between the ENZ response of ITO and the plasmonic modes of Ag produces hybrid ENZ–plasmon resonances, whereas exciton–polaritons are light–matter hybrid eigenstates. Characterization reveals stronger dispersion and higher phase sensitivity for transverse magnetic polarization compared to transverse electric polarization (Appendix C). Introducing time-dependent permittivities makes the photonic Hamiltonian explicitly time-dependent, allowing ultrafast material dynamics to reshape its eigenvalue spectrum in real time (Appendix B). The temporal deformation drives the motion of exceptional points and enables quantized topological transitions on femtosecond–picosecond timescales. The temporal response of the multilayer is modeled using a time-dependent Drude–Lorentz permittivity for ITO, which accounts for both free-carrier dynamics and interband transitions, $\varepsilon_{ITO}(\omega,t) = \varepsilon_\infty - \frac{\omega_p^2(t)}{\omega^2+i\gamma(t)\omega} + \sum_j \frac{f_j\omega_j^2}{\omega_j^2-\omega^2-i\Gamma_j\omega}$. Here, $\varepsilon_\infty$ is the high-frequency dielectric constant, while $\omega_p(t)$ and $\gamma(t)$ denote the time-dependent plasma frequency and damping factor. Carrier relaxation in ITO follows a single-exponential decay with characteristic time $\tau_{ITO}$, $\omega_p(t) = \omega_p(0)e^{-t/\tau_{ITO}}$, corresponding to a carrier density $n(t) = n_0 e^{-t/\tau_{ITO}}$, where $\omega_p(t) = \sqrt{n(t)e^2/(\varepsilon_0 m^*)}$. SiO$_2$ is modeled with a weak, phenomenological modulation of its refractive index to include secondary nonlinear effects, $\varepsilon_{SiO_2}(\omega,t) = \varepsilon_{SiO_2}^0 + \Delta\varepsilon_{SiO_2}(t)$, where $\varepsilon_{SiO_2}^0$ is the static permittivity, and $\Delta\varepsilon_{SiO_2}(t) \propto e^{-t/\tau_{SiO2}}$ represents slow, thermally driven relaxation due to Kerr-type nonlinearities. Although weaker than the ITO contribution, this term captures secondary nonlinear effects that slightly influence the resonance position and modal confinement, improving the accuracy of the simulated optical response. The spectral reflectance and field distributions indicate the ENZ condition near $\lambda \approx 510$ nm, where $Re[\varepsilon_{ITO}] \approx 0$ (Figs. 1b). The real and imaginary parts of $\varepsilon_{ITO}$, together with the corresponding field enhancement, demonstrate strongly dispersive phase behavior and intense electric-field confinement at the ENZ wavelength (Figs. 1c). The characteristics are essential for interpreting the temporal evolution of the multilayer's optical properties. Angle–wavelength reflectance maps at $t = 0.2$ ps and $t = 10$ ps (Figs. 1d,k) show a distinct reflectance minimum. At $t = 0.2$ ps, the low-reflectance region spans $\lambda \approx 460$–495 nm, with a local minimum near 490 nm in the fixed-angle cross-section at $\vartheta = 45.6°$ (Figs. 1h). As relaxation progresses, a second low-reflectance band appears at longer wavelengths (lower photon energy), concurrent with the redshift of the primary resonance. The feature arises from the altered permittivity landscape of ITO, which modifies the coupling between plasmonic and ENZ modes. The emergence and evolution of this additional band indicate a reorganization of the system's eigenmodes, consistent with a change in the underlying spectral topology as the non-Hermitian parameters evolve in time. Moreover, by $t = 10$ ps, the minimum redshifts to $\lambda \approx 520$–525 nm and broadens (Figs. 1g,k), consistent with carrier relaxation in ITO Appendix A). The $2\pi$-modulo phase $\phi = \arg[r(\lambda,\theta,t)]$ represents the principal-value phase constrained to $(-\pi,\pi]$ (Figs. 1e,h). Apparent discontinuities correspond to $\pm 2\pi$ jumps at reflectance minima where $|r| \to 0$, identifying points of strong modal coupling. The unwrapped phase $\tilde{\phi}$ removes the jumps, producing a continuous phase map over the $(\lambda,\theta)$ domain (Figs. 1f,i). Between 0.2 ps and 10 ps, the low-reflectance region redshifts from $\approx 490$ nm to $\approx 525$ nm and broadens, while the associated phase discontinuities shift accordingly. The cumulative-phase slope, visible in fixed-angle cross-sections (Figs. 1j,k), decreases with delay, reflecting a transient change in dispersion. The simultaneous evolution of amplitude and phase reveals that the hybrid ENZ–plasmonic resonance remains active while its spectral position and phase topology reorganize in time. The correlated shift of reflectance minima and phase discontinuities indicates that the underlying permittivity—and hence the eigenmode structure-changes continuously during carrier relaxation in ITO, linking the material's ultrafast response to the temporal evolution of the photonic phase.

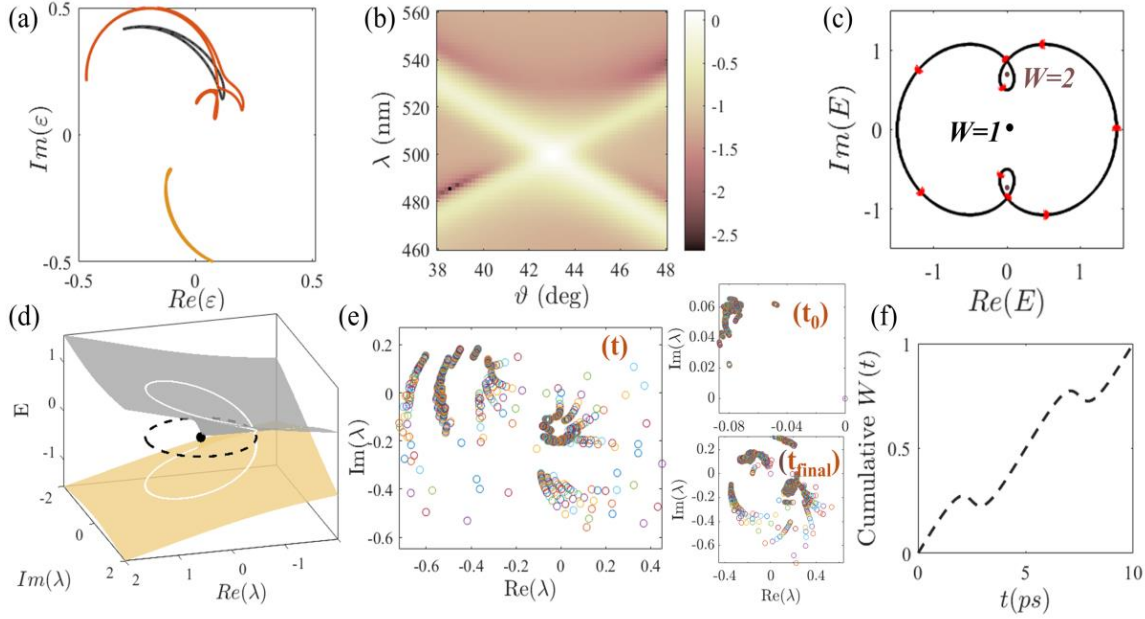

*Figure 2. Ultrafast evolution of non-Hermitian topology. (a) Permittivity trajectories of ITO and $SiO_2$ in the complex plane. (b) $\log|\Delta|$ and X-shaped avoided crossing. (c) Topological vortices ($W = 1,2$) in the $(\lambda, \vartheta)$ plane corresponding to exceptional points. (d) Riemann surface of $\Lambda_{1,2}$. (e) Temporal evolution of exceptional point positions as vortex trajectories during relaxation. (f) Cumulative winding number $W(t)$*

The time-dependent non-Hermitian operator $H(\lambda, \vartheta, t)$, its complex eigenvalues $\Lambda_{1,2}(\lambda, \vartheta, t)$, and the discriminant $\Delta(\lambda, \vartheta, t) = [\Lambda_1(\lambda, \vartheta, t) - \Lambda_2(\lambda, \vartheta, t)]^2$ link the multilayer's material dynamics to its topological evolution (Appendix B). Because the ITO/$SiO_2$/Ag multilayer supports both absorption and radiative leakage, $H$ is intrinsically non-Hermitian, and its eigenvalues are complex—the real parts describe resonance energies, while imaginary parts represents linewidths (Appendix D). The two coupled eigenmodes correspond to the plasmonic mode at the Ag/ITO interface and the ENZ-guided mode in the ITO–$SiO_2$ region. Temporal variations of the ITO permittivity continuously deform $H(t)$, modulating the hybridization strength between the modes and displacing their exceptional points in the $(\lambda, \vartheta)$ plane. Fig. 2a shows the trajectories of the complex permittivity of ITO and $SiO_2$ in the complex plane. Each curve traces how photoexcitation and carrier relaxation transiently modify both the dispersive ($Re[\varepsilon]$) and absorptive ($Im[\varepsilon]$) components. Since the elements of $H(t)$ depend explicitly on these permittivities, the motion of $\varepsilon(t)$ represents a continuous deformation of the Hamiltonian as the system evolves in time. The corresponding change in the spectral geometry is captured by the discriminant magnitude, $\log|\Delta|$, which displays the characteristic X-shaped avoided crossing between the two hybrid modes (Fig. 2b). The phase of the discriminant, $\arg[\Delta]$, defines the topological phase field in the $(\lambda, \vartheta)$ domain: its zeros mark exceptional points where the two hybrid branches coalesce (Fig. 2c). Each singularity carries a winding number $W$, given by the $2\pi$ phase circulation of $\arg[\Delta]$ around the degeneracy. Singularities with $W = 1$ represent first-order exceptional points formed by the coalescence of the plasmonic and ENZ modes. Under stronger coupling conditions, when two such degeneracies merge or are simultaneously enclosed, the total charge increases to $W = 2$. The presence of the two winding numbers reflects the presence of two exceptional points supported by the multilayer geometry and the evolving ITO permittivity.

The Riemann surface of the complex eigenvalues, reconstructed from both branches $\Lambda_{1,2}(\lambda, \vartheta, t)$, exhibits the expected square-root topology of non-Hermitian degeneracies (Fig. 2d). The two sheets meet at each exceptional point, and encircling a singularity exchanges the modes between sheets, confirming the nontrivial topology of the coupled system. The temporal evolution of the singularities is shown in Fig. 2e. As permittivity relaxes, the exceptional points drift across the parameter space, and the eigenvalues trace vortex-like loops in the complex plane. The emergence of a loop corresponds to vortex nucleation ($\Delta W = +1$), while its disappearance would indicate annihilation ($\Delta W = -1$); in the present observation, only the nucleation event ($0 \to 1$) occurs, consistent with the time window of carrier relaxation. The eigenvalues approach but do not fully coalesce, showing that the system passes near the exceptional point without complete modal collapse. The partial separation preserves spectral distinction and allows the topological evolution to be temporally resolved. The motion of the eigenvalues determines the instantaneous resonance positions and linewidths, such that the vortex formation in the eigenvalue plane is directly manifested in the transient reflectance response. Finally, the cumulative winding number $W(t) = \frac{1}{2\pi} \oint_C d \arg[\Delta(\lambda, \vartheta, t)]$ (Fig. 2f) quantifies the time-dependent topology. Because $W(t)$ is integer-valued, it changes only when an exceptional point crosses the contour $C$. The increased from 0 to 1 signifies the creation of a single topological vortex during carrier relaxation in ITO. As such, the results establish a causal chain linking ultrafast permittivity modulation to non-Hermitian logical evoltopoution in the multilayer: temporal change of $\varepsilon_{ITO} \to$ deformation of $H(t) \to$ motion of exceptional points $\to$ discrete change in $W(t)$. The avoided crossings, phase vortices, Riemann-surface topology, and eigenvalue trajectories provide complementary evidence of similar mechanisms, where the exceptional point acts as both a spectral singularity and a source of quantized phase winding in the hybrid ENZ–plasmonic system.

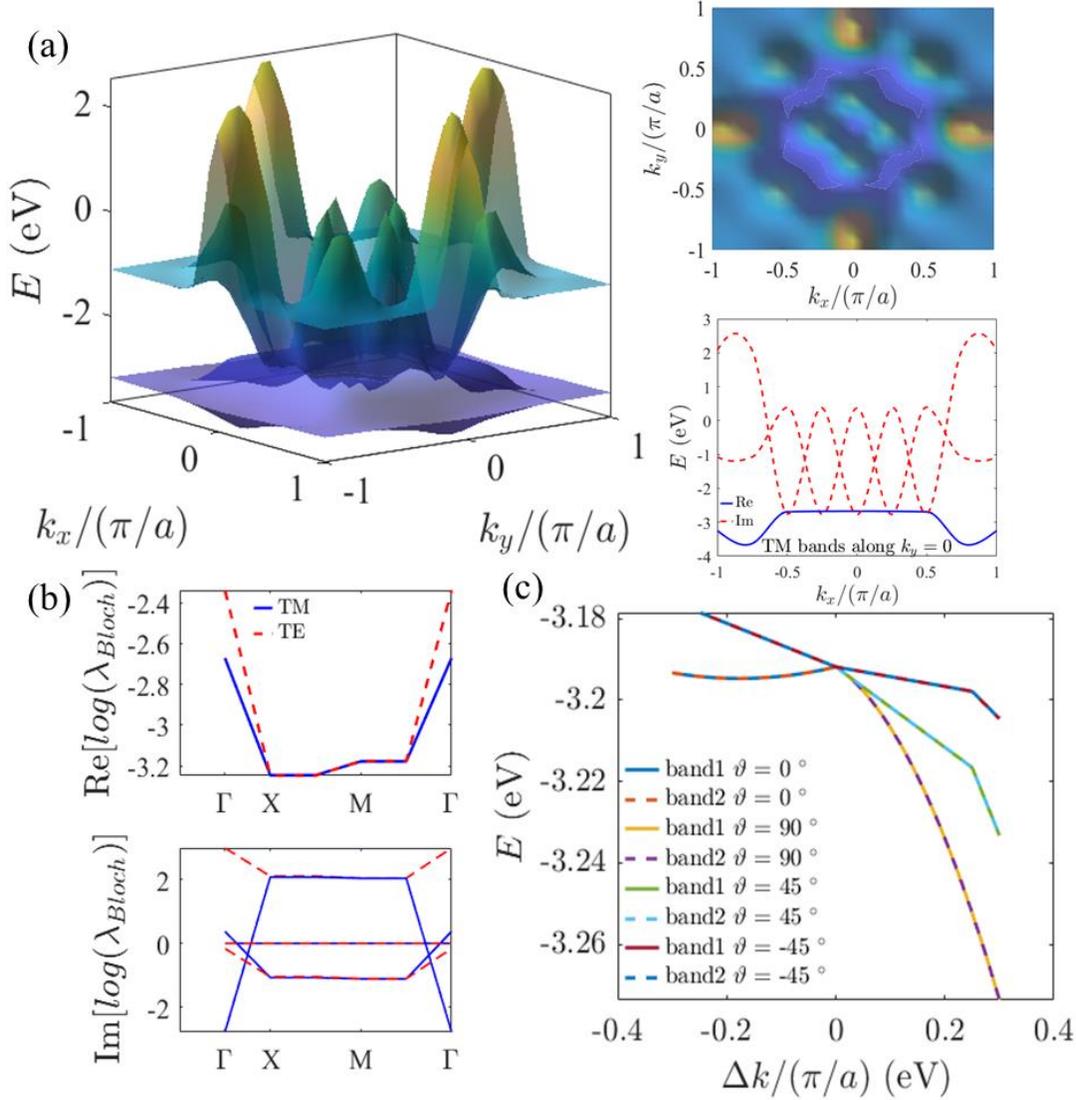

*Figure 3. Ultrafast modulation of photonic band topology. (a) Quasi-energy band surface at t=0.5 ps. Insets: details on the sideband rings and transient avoided crossings. (b) Complex Bloch eigenvalues $log(\lambda_{Bloch})$ along $\Gamma$–$X$–$M$–$\Gamma$: transverse magnetic modes (solid), transverse electric modes (dashed). (c) transverse magnetic dispersion along $k_y = 0$ exhibits a linear crossing near 1.9 eV, a Type-I Dirac point arising from plasmon–ENZ coupling.*

Ultrafast modulation of the multilayer permittivity dynamically reshapes its photonic band topology in real time. The calculated complex band dispersion (Fig. 3) illustrates how time-dependent modulation induces avoided crossings and mediates the transfer of topological charge. In panel (a), the three-dimensional band surface shows that the clean stop band of 0.9–1.3 eV observed in the static structure (Appendix E) evolves into a sequence of avoided crossings and transient gaps extending across the 1.5–2.5 eV range. The anticrossings originate from Floquet-like hybridization driven by the periodic modulation of the permittivity, which mixes the fundamental photonic band with its temporally shifted counterparts. Because at $t = 0.5$ ps, the phase vortex in Fig. 2 approaches contour $C$, the band structure is frozen at that instant. The interaction between the components opens dynamic gaps at their intersections in $(\mathbf{k}, E)$ space and redistributes spectral weight across the hybridized branches. Moreover, the strongest hybridization occurs during the modulation interval where the phase vortex in Fig. 2 approaches the contour $C$, linking the onset of sideband coupling to the topological charge transfer responsible for the discrete change in $W(t)$. The right panel of (a) shows the corresponding momentum-space intensity distribution. Under temporal modulation, concentric Dirac-like rings emerge—features absent in the static spectrum—that mark the regions where the driven field couples the primary photonic band to its modulation sidebands. The rings provide a direct view of how ultrafast permittivity modulation transfers energy and phase between temporally shifted Bloch modes. The dispersion along $k_y = 0$ in the lower-right panel further reveals a sequence of avoided crossings and oscillatory energy shifts, indicating strong temporal mixing between the fundamental band and its modulation-induced harmonics. Rather than producing sharp Dirac intersections, the hybridized branches display non-Hermitian broadening and complex-valued coupling, which open finite dynamic gaps and enhance attenuation near resonance. The Floquet-like band maps the dispersion onto an extended quasi-energy Brillouin zone, where temporally shifted harmonics overlap and hybridize. The corresponding imaginary-band component (red dashed curve) exhibits a periodic modulation, identifying alternating regions of dissipation and amplification governed by the time-dependent permittivity of ITO. The transient conical features mediate the topological charge transfer observed in Fig. 2, providing the microscopic mechanism for the temporal redistribution of Berry curvature during ultrafast modulation. It is worth noting that the Floquet–Dirac features appear exclusively under transverse magnetic polarization, consistent with the strong longitudinal-field coupling to the ENZ response of ITO. The transverse electric comparison (Appendix E, Fig. 8) shows no equivalent linear crossings or sideband mixing, confirming that the Floquet–Dirac behavior is intrinsically polarization selective and tied to the non-Hermitian longitudinal-field dynamics of transverse magnetic modes. The complex Bloch eigenvalues (Fig. 3b), $\log(\lambda_{\text{Bloch}}) = ik_z D$, plotted along $\Gamma$–$X$–$M$–$\Gamma$, highlight the non-Hermitian character of the temporally modulated structure. The imaginary part, $k_{z'}D$, corresponds to phase accumulation per unit cell, while the real part, $-k_{z''}$, quantifies attenuation. Around ENZ-assisted resonances, the TM-polarized bands exhibit rapid oscillations in $\text{Im}[\log(\lambda)]$ and strong negative shifts in $\text{Re}[\log(\lambda)]$, marking regions where longitudinal fields couple efficiently to the time-dependent carrier plasma, inducing enhanced loss and phase dispersion. In this regime, the eigenvalues trace closed trajectories in the complex plane rather than lying on a single branch, identifying the onset of exceptional-point dynamics driven by temporal modulation. The formation of these loops reflects the interplay between ENZ-induced dissipation and quasi-Floquet hybridization, which jointly deform the complex band manifold. Compared to the static case (Appendix E), temporal modulation therefore not only replicates the bands through sideband coupling but also introduces an additional non-Hermitian degree of freedom that continuously rotates the eigenvalue phases during the modulation cycle. The contrast with the transverse electric polarization shows that the latter lacks a longitudinal electric-field component, experience negligible coupling to the ENZ response and remain nearly Hermitian. Detailed dispersion in Fig. 3c displays a cascade of avoided crossings and oscillatory energy shifts along $k_y = 0$, confirming strong temporal mixing between the fundamental band and its harmonics. Near 1.9 eV, the two transverse magnetic branches intersect linearly, forming a Type-I Dirac point that originates from the coupling between the plasmonic and ENZ-like modes. Rather than remaining sharp, this intersection broadens under modulation, producing finite dynamic gaps and non-Hermitian attenuation linked to the complex-valued ITO permittivity. The transient conical features mediate the topological charge exchange seen in Fig. 2, providing the microscopic mechanism for the ultrafast redistribution of Berry curvature. The Floquet–Dirac response appears exclusively under transverse magnetic excitation, consistent with the longitudinal-field coupling to the ENZ response of ITO. The transverse electric comparison (Appendix 6, Fig. 8) shows no equivalent linear crossings or sideband mixing, confirming that the Floquet–Dirac behavior is intrinsically polarization-selective and tied to the non-Hermitian longitudinal-field dynamics of transverse magnetic modes. The behavior originates from the time-dependent Drude–Lorentz response of ITO, where photoexcited carrier relaxation alters both $\text{Re}[\varepsilon_{\text{ITO}}]$ and $\text{Im}[\varepsilon_{\text{ITO}}]$ on a picosecond timescale. The transient decrease in plasma frequency and damping rate modifies hybrid plasmon–ENZ coupling strength, shifting the eigenvalue landscape of the multilayer and driving the observed evolution of band gaps, Dirac intersections, and topological charge. Thus, the Floquet–Dirac dynamics directly trace the ultrafast permittivity trajectory of ITO and quantify how carrier-induced non-Hermiticity governs real-time topological reconfiguration. In conclusion, the study establishes a direct physical link between ultrafast carrier relaxation in ENZ media and the real-time evolution of non-Hermitian topology. By modeling the time-dependent permittivity of ITO within a hybrid ENZ–plasmonic multilayer, we resolve the motion of exceptional points and discrete winding-number shifts induced by picosecond modulation. Temporal variations in dispersion and dissipation alone can trigger topological phase transitions without structural alteration, unifying non-Hermitian singularities and Floquet hybridization as dual manifestations of the same ultrafast ENZ-driven dynamics.

# Appendix A

# Numerical Framework

The numerical procedure begins with the computation of the complex reflection coefficient $r(\lambda, \vartheta, t)$ and, more generally, the eigenvalues $s_n(\mathbf{p}, t)$ of the scattering matrix of the multilayer system. For static studies, the parameter vector is $\mathbf{p} = (\lambda, \vartheta)$, where $\lambda$ is the wavelength and $\vartheta$ the incidence angle. For driven studies, we use $\mathbf{p} = (\Re[\varepsilon], \Im[\varepsilon])$, where $\epsilon$ is the complex dielectric permittivity of ITO. Eigenvalues evolve continuously in the complex plane as the parameters vary, and their trajectories encode the presence of singularities. RCWA simulations were performed using a plane-wave basis with $N = \pm N_{\text{harm}}$ diffraction orders, converged at $N_{\text{harm}} = 3$. Periodic boundary conditions match the microrod lattice, which has period $\Lambda = 100$ nm and rod diameter $D = 50$ nm. The Ag microrod layer is modeled either as a full periodic grating or, when appropriate, using an effective-medium approximation validated against the full structure. Transverse magnetic and transverse electric polarizations are defined with respect to the plane of incidence, with transverse magnetic corresponding to a magnetic field parallel to the microrod axis. Convergence of reflectance and phase spectra was verified to within 1% over the full wavelength–angle domain (460 nm $\leq \lambda \leq$ 780 nm, 43.8° $\leq \vartheta \leq$ 45.6°). To characterize singularities, we compute winding numbers derived from the cumulative, unwrapped phase:

$$z_k = s_n(\mathbf{p}_k, t) - s^*,$$

$$\Phi = \Im \sum_{k=1}^{K} \log\left(\frac{z_{k+1}}{z_k}\right),$$

$$w_n(t; C) = \text{round}\left(\frac{\Phi}{2\pi}\right),$$

where $\{\mathbf{p}_k\}_{k=1}^{K}$ is a discrete closed loop in parameter space sampled finely enough to capture singularities, and $s^* = 0$ is the reference point in the complex plane. The unwrapping ensures that the cumulative phase is continuous along the loop, yielding integer winding numbers invariant under continuous parameter variation. Exceptional points are identified by monitoring eigenvalue coalescence and the ill-conditioning of the eigenvector matrix:

$$\min_{i \neq j} |s_i - s_j| < \varepsilon,$$

$$\kappa(V) = \| V \| \| V^{-1} \| > \kappa_{\text{thresh}},$$

where $V$ is the matrix of eigenvectors and $\varepsilon$ and $\kappa_{\text{thresh}}$ are numerical thresholds. Additional discriminant checks and parameter-continuation methods allow the reconstruction of smooth exceptional point trajectories $\mathbf{p}_{\text{EP}}(t)$.

*Drude parameters used for ITO and $SiO_2$.*

| Material | Symbol | Value | Units |
|---|---|---|---|
| ITO | $\varepsilon_\infty$ | 4.0 | — |
|  | $\omega_p(0)$ | $9.6 \times 10^{15}$ | $s^{-1}$ |

| Material | Symbol | Value | Units |
|---|---|---|---|
| | $\gamma(0)$ | $2.0 \times 10^{14}$ | $s^{-1}$ |
| SiO$_2$ | $\varepsilon_{SiO_2}^0$ | 2.1 | — |
| | $\tau_{SiO_2}$ | $1.0 \times 10^{-12}$ | s |

When a loop $C_j$ encloses a single exceptional point, the winding number changes by exactly $\Delta w_n^{(j)} = \pm 1$; if multiple exceptional numbers are enclosed, $\Delta w_n$ is the sum over all exceptional points. When an exceptional point exits the loop, the winding number returns to its original value. This dynamic, quantized behavior is a hallmark of non-Hermitian Floquet topology. Validation is provided using test cases such as circular loops around a known singularity, where the cumulative phase increases exactly by $2\pi$, giving $w = 1$ independent of time. The tests confirm the stability of the winding-number extraction against discretization, phase discontinuities, and ultrafast modulation. **Table S1** shows the parameters used for the numerical analysis.

# Appendix B

## Explicit Hamiltonian Model

The non-Hermitian Hamiltonian $H(t)$ describing the coupled optical modes of the ITO/SiO$_2$ heterostructure is constructed from the instantaneous dielectric functions obtained in the pump–probe measurements. A minimal two-mode representation is

$$H(t) = \begin{pmatrix} \omega_1(\varepsilon_{ITO}(t)) - i\gamma_1(\varepsilon_{ITO}(t)) & \kappa(\varepsilon_{ITO}(t), \varepsilon_{SiO_2}(t)) \\ (6pt]\kappa(\varepsilon_{ITO}(t), \varepsilon_{SiO_2}(t)) & \omega_2(\varepsilon_{SiO_2}(t)) - i\gamma_2(\varepsilon_{SiO_2}(t)) \end{pmatrix}.$$

Here, $\omega_j$ are the real resonant frequencies determined by $\text{Re}(\varepsilon_j)$, $\gamma_j$ are the effective loss rates proportional to $\text{Im}(\varepsilon_j)$, and $\kappa$ represents the coupling strength between the two modes. The time dependence of $\varepsilon_j(t)$ extracted from Fig. 2c,d directly modulates all matrix elements of $H(t)$, driving its continuous deformation. The instantaneous eigenvalues $E_\pm(t)$ of this matrix yield the trajectories shown in Fig. 2e. Exceptional points correspond to the condition

$$(\omega_1 - \omega_2 - i(\gamma_1 - \gamma_2))^2 + 4\kappa^2 = 0,$$

and their temporal motion in parameter space reflects the dynamic balance between dispersion, absorption, and coupling controlled by the evolving permittivities.

# Appendix C

# Static Characterization and Baseline Response

A realistic description of the multilayer requires incorporating the static modal response and field distribution under both transverse magnetic and electric polarizations. Rigorous coupled-wave analysis simulations were performed with the multilayer thicknesses specified in **Table S2**.

*Multilayer layout and layer thickness[a]*

| Layer | Material | Thickness | Refractive index |
|---|---|---|---|
| 1 | ITO | 80–100 nm | $1.83 + 0.0031\,i$ |
| 2 | $SiO_2$ | 500 $\mu$m | 1.52 |
| 3 | (3-mercaptopropyl) trimethoxysilane | 0.1 nm | 1.44 |
| 4 | Ag | 50–70 nm | $0.052 + 3.9\,i$ |
| 5 | Rhodamine | 5–8 nm | $1.40 + 0.024\,i$ |

[a]Approximate thicknesses and refractive indices.

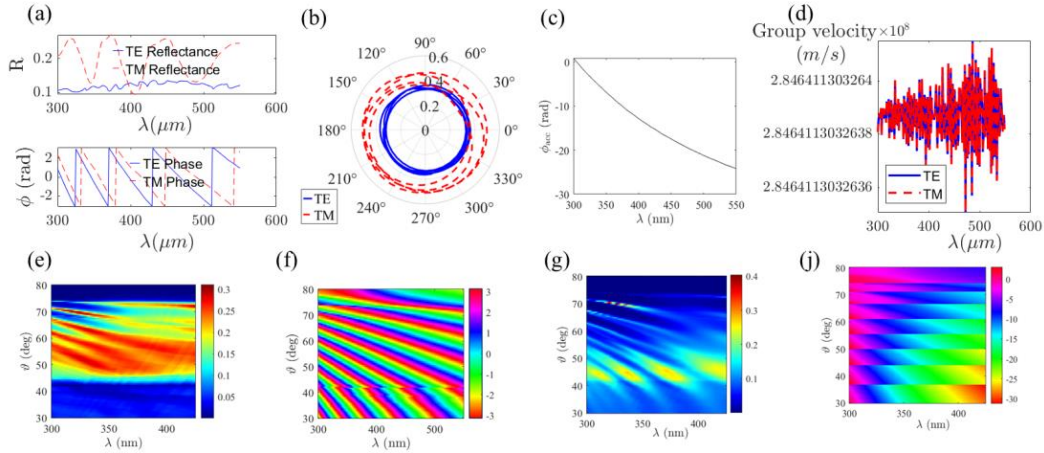

*Figure 4. Spectral and angular analysis of the steady-state reflectance and phase response of the multilayer system with Drude-dispersive media. (a) Normal-incidence transverse electric and transverse magnetic reflectance (top) and phase (bottom). (b) Complex reflection coefficient in the Argand plane. (c) cumulative has accumulation. (d) Group velocity. (e) transverse electric reflectance map. (f) transverse electric phase map. (g) transverse magnetic reflectance map. (j) transverse magnetic phase map.*

The static optical response of the multilayer system is summarized in Fig. 4, combining reflectance, phase, and angular maps into a unified modal picture. At normal incidence, the transverse electric and magnetic reflectance spectra $R(\lambda)$ (Fig. 4a, top) remain below 0.3 across 300–600 nm, indicating the absence of strong Fabry–Pérot or plasmonic resonances. Transverse magnetic reflectance shows quasi-periodic oscillations with a free spectral range $\Delta\lambda \approx 650\ nm$ ($n_{\text{eff}}d \approx 1.8\ \mu m$). In the 300–400 nm region the narrow to $\Delta\lambda \approx 80$ nm, while transverse electric features stay broad and weak. The phase $\phi(\lambda) = \arg[r(\lambda)]$ (Fig. 4a, bottom) shows smoother transverse electric evolution but a steeper transverse magnetic slope, $d\phi/d\lambda \approx 0.07$ rad/nm versus 0.04 rad/nm, indicating stronger transverse magnetic dispersion and group delay. Argand-plane trajectories (Fig. 4b) enclose larger loops for transverse magnetic (~ 0.45 area, $0.3\pi$ phase) than transverse electric (~ 0.15, $0.1\pi$), consistent with ~ 15% deeper transverse magnetic field penetration. The cumulative phase (Fig. 4c) tracks reflectance fringes, confirming modal coherence. From the phase slopes, the group velocity $v_g(\lambda)$ (Fig. 4d) is ~ $0.92c$ off-resonance, decreasing to $0.78c$ (transverse electric) and $0.65c$ (transverse magnetic) near dispersive regions. Angularly resolved maps show weak transverse electric dispersion (~ 0.25 nm/deg) and stronger transverse magnetic shifts (~ 0.4 nm/deg) with sharper phase gradients up to $0.5\pi$ over 400 nm. Overall, transverse electric modes remain weakly dispersive and nearly isotropic, whereas transverse magnetic modes exhibit stronger confinement, dispersion, and slow-light effects. This polarization asymmetry defines the system's static optical fingerprint, providing a baseline for dynamic tuning and topological analysis. Angularly resolved reflectance and phase maps reveal the momentum dependence of these modes. The transverse electric reflectance $R(\lambda, \theta)$ (Fig. 4e) exhibits dispersive branches that blue-shift at ~ 0.25 nm/deg, consistent with weak leaky-mode coupling. Its phase map (Fig. 4f) shows moderate gradients up to $0.3\pi$ over $\Delta\lambda \approx 5$ nm, aligned with reflectance minima. transverse magnetic reflectance (Fig. 4g) is more strongly dispersive, with denser features shifting at ~ 0.4 nm/deg, characteristic of surface- or quasi-surface plasmonic modes. The corresponding transverse magnetic phase map (Fig. 4j) exhibits even sharper gradients, up to $0.5\pi$ over 4 nm, indicating tighter confinement and higher sensitivity to in-plane momentum. Altogether, the spectral (Figs. 4a,c), phase (Figs. 4a,b,f,j), group velocity (Fig. 4d), and angular (Figs. 4e–j) analyses provide a consistent picture. transverse electric modes remain weakly dispersive and nearly isotropic, while transverse magnetic modes show stronger dispersion, enhanced slow-light effects, larger phase accumulation, and sharper angular sensitivity. This polarization asymmetry forms a quantitative static fingerprint of the multilayer system, serving as a baseline against which dynamic tuning, topological winding, and exceptional-point behavior can be compared.

The time-resolved reflectance and phase response of the multilayer with time-dependent optical properties are presented in Figure 5. At $t = 0$ (Figure 5), the multilayer exhibits the maximum carrier density, establishing the baseline resonance and phase response that serve as the reference for all subsequent time-dependent analysis. The reflectance amplitude map $r(\lambda, \theta)$ shows a shallow minimum within $\lambda \sim 480$–510 nm at incidence angles around $\theta \sim 45°$ (panel a). The feature is consistent with the spectral region expected for the hybrid plasmonic mode sustained by the ITO/SiO$_2$/Ag interface, although its signature is not sharply resolved in this map. The modulo-$2\pi$ phase $\phi = \arg(r)$ map (Figure 5b) shows abrupt discontinuities that occur exactly where the reflectance reaches its minimum. The jumps in phase correspond to phase discontinuities in the optical response, arising from strong modal dispersion and interference between different optical paths.

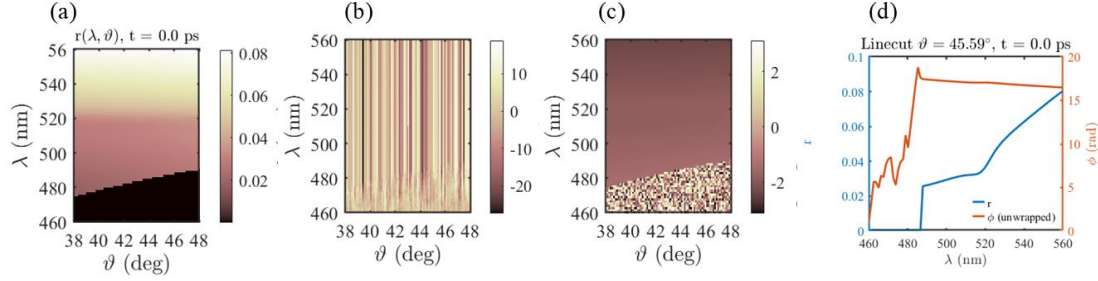

*Figure 5. Static optical response of the multilayer at $t = 0$. (a) Reflectance map $R(\lambda, \theta)$ showing the angular–wavelength dispersion of the hybrid plasmonic mode. (b) Modulo-$2\pi$ phase $\phi = arg(r)$. (c) Cumulative phase $\tilde{\phi} = unwrap[arg(r)]$. (d) cross sections of $r$ and $\phi$.*

At shorter wavelengths, residual numerical noise becomes visible when the reflectance amplitude is very low; these artifacts do not affect the physical interpretation but manifest as small fluctuations in the phase map. When the phase is cumulative $\tilde{\phi} = unwrap[arg(r)]$ (Figure 5c), global continuity is restored; the phase discontinuities associated with $|r| \to 0$ appear as steep ramps and extended plateaux. The fixed-angle cross-section at $\vartheta = 45.6°$ (panel d) makes the correspondence clear: the reflectance dip is accompanied by a sharp increase in the cumulative phase.

# Appendix D

# Topology in dynamic conditions

To clarify the link between phase singularities and the discriminant formalism, we analyzed the steady-state reflectance and phase response of the ITO/SiO$_2$/Ag multilayer before and after optical modulation. Figures 6(a–b) show the phase vortex in the $(\lambda, \vartheta)$ plane, revealing the position of the singularity and how it shifts under modulation. The corresponding discriminant maps, $|\Delta|$, shown on a logarithmic scale in Figures 6(c–d), display sharp X-shaped features that follow the vortex displacement, providing a direct representation of the exceptional-point topology in the scattering response.

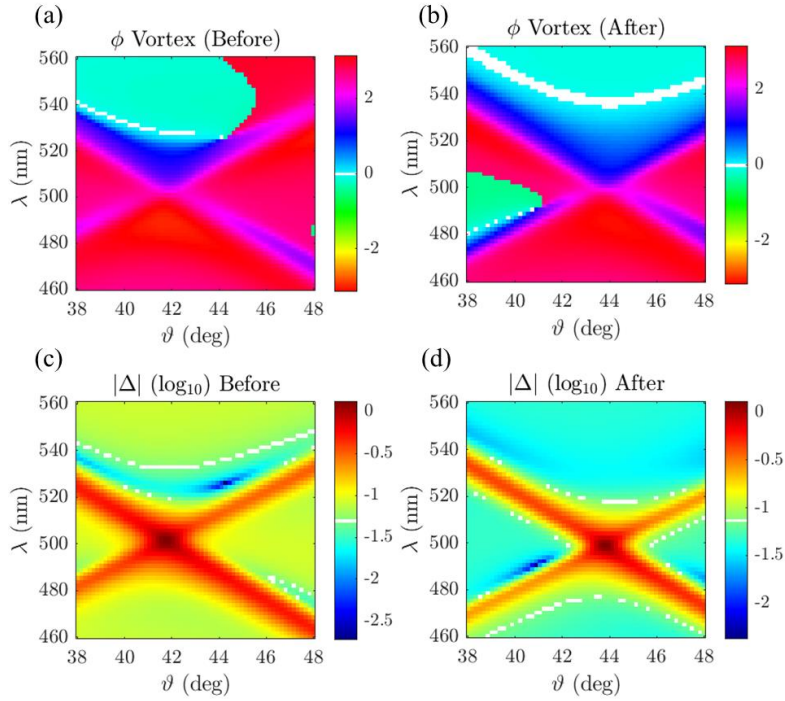

*Figure 6. Spectral and angular analysis of the steady-state response of the indium tin oxide/SiO$_2$ multilayer. (a–b) Phase vortex maps in the $(\lambda, \vartheta)$ plane before and after modulation, showing the displacement of the singularity associated with exceptional-point crossing. (c–d) Discriminant magnitude $|\Delta|$ on a logarithmic scale, highlighting X-shaped features that coincide with the vortex displacement and reveal the underlying non-Hermitian topology. These steady-state diagnostics provide scattering-level evidence of exceptional-point dynamics and complement the band structure analysis presented in Fig. 3.*

# Appendix E

# Static Complex Band Analysis

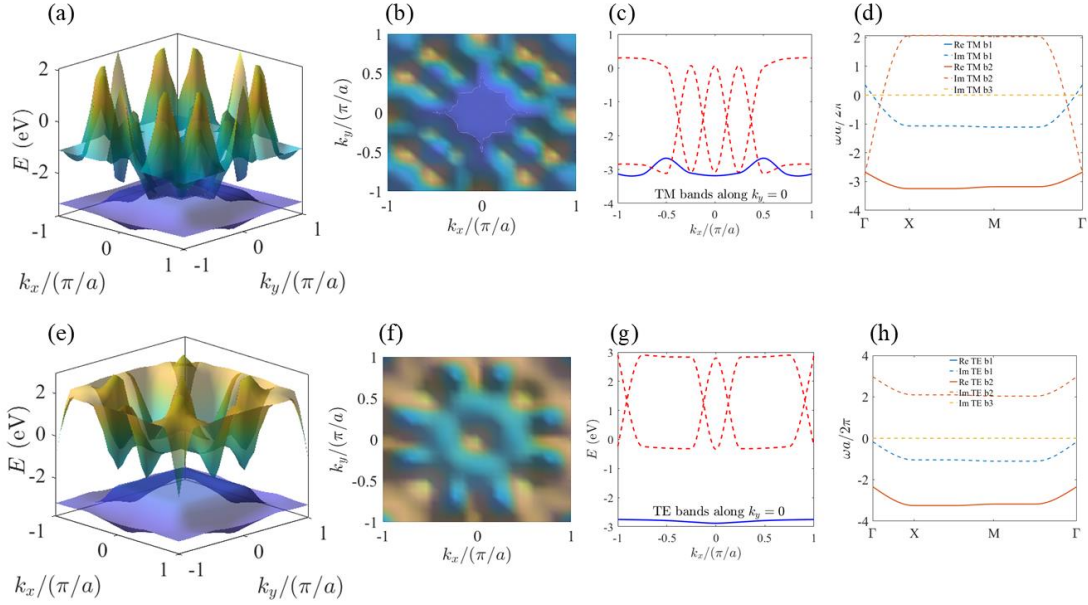

*Figure 7. Band structures of the multilayer under static conditions for transverse magnetic (a–d) and transverse electric (e–h) polarizations. Panels (a,e) show the 3D energy surfaces $E(k_x, k_y)$; (b,f) the corresponding top-down reciprocal-space projections; (c,g) the dispersions along $k_y = 0$; and (d,h) the band diagrams along high-symmetry paths (Γ–X–M). Transverse magnetic modes exhibit a pronounced stop band around $\omega a/2\pi \approx 1.0$, strong anisotropy, and mode splitting, while transverse electric modes show nearly isotropic dispersion with only a shallow pseudogap.*

Ultrafast modulation of photonic band topology. (a) Quasi-energy band surface at t=0.5 ps. Insets: details on the sideband rings and transient avoided crossings. (b) Complex Bloch eigenvalues $\log(\lambda_{\text{Bloch}})$ along Γ–X–M–Γ: transverse magnetic modes (solid), transverse electric modes (dashed). (c) transverse magnetic dispersion along $k_y = 0$ exhibits a linear crossing near 1.9 eV, a Type-I Dirac point arising from plasmon–ENZ coupling.

For transverse magnetic polarization (Figs. 7a–d), the 3D energy surface $E(k_x, k_y)$ in Fig. 7a exhibits a pronounced stop band centered near $\omega a/2\pi \approx 1.0$ (0.9–1.3 eV) with a width of ~ 0.4 eV. Its top-down reciprocal-space projection (Fig. 7b) reveals strong anisotropy, with elongated valleys along the Γ → X direction indicative of direction-dependent group velocities and reduced transmission away from the zone center. The $k_y = 0$ dispersion cut (Fig. 7c) shows multiple transverse magnetic branches and clear mode splitting: the lowest transverse magnetic branch is relatively flat (estimated $v_g \sim 0.2c$), whereas higher-order branches ($E \approx 2.5$–3 eV) are gapped and strongly attenuated. The high-symmetry band diagram plotted as $\omega a/2\pi$ across Γ − X − M (Fig. 7d) confirms a well-defined transverse magnetic bandgap near $\omega a/2\pi \approx 1.0$, consistent with suppressed propagating states across the Brillouin zone. By contrast, for transverse electric polarization (Figs. 7e–h), the 3D surface (Fig. 7e) shows only a shallow pseudogap around $\omega a/2\pi \approx 1.0$ with width $\sim 0.1$ eV. The corresponding top-view map (Fig. 7f) is nearly isotropic, indicating largely uniform in-plane transport. The $k_y = 0$ cut for transverse electric (Fig. 7g) displays continuous dispersion below $E \approx 2$ eV with minimal splitting and attenuation confined to higher bands, and the $\Gamma - X - M$ band plot (Fig. 7h) corroborates the absence of a substantial transverse electric gap and the dominance of propagating modes. Together, these panels demonstrate a pronounced polarization asymmetry: transverse magnetic modes show wider gaps, stronger attenuation, and higher anisotropy, while transverse electric modes are largely continuous and isotropic. The normalization of momentum to $k/(\pi/a)$ permits direct comparison across the multilayer design.

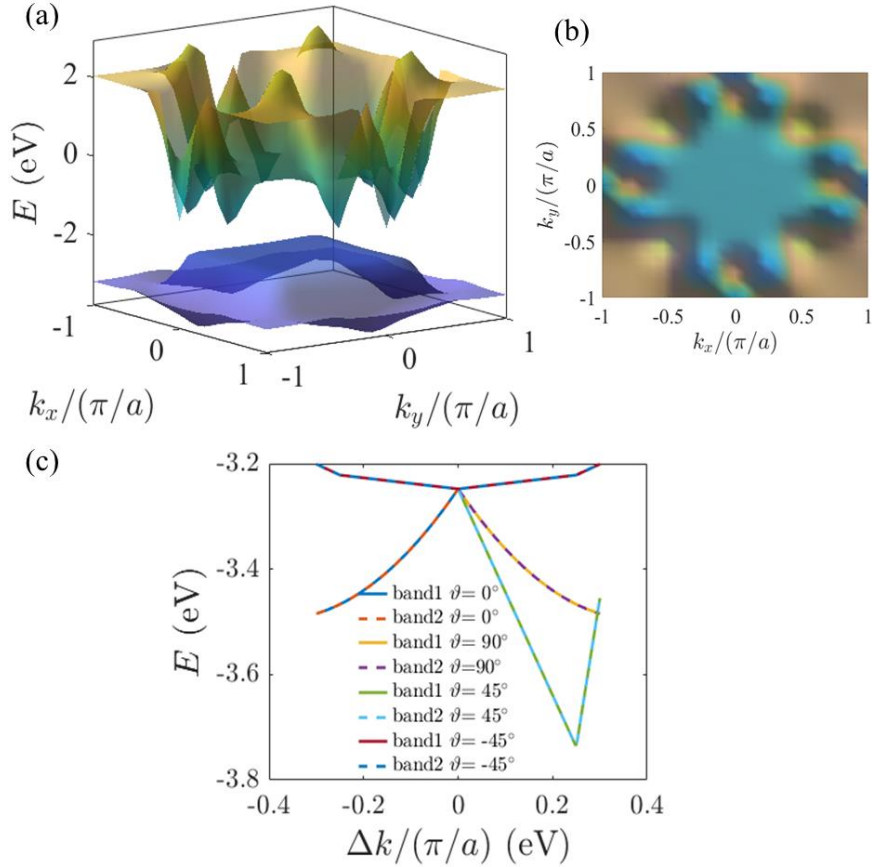

*Figure 8. Dynamic band structure of the multilayer for transverse electric (TE) polarization under time-dependent permittivity modulation. (a) 3D energy surface $E(k_x, k_y)$ and corresponding top-down reciprocal-space projection reveal band reshaping and weak anisotropy induced by temporal modulation. (b) Angularly resolved dispersions plotted as a function of $\Delta k/(\pi/a)$ for different propagation angles $\phi$ show direction-dependent band splitting and frequency shifts near the crossing point, consistent with a slightly tilted type-I Dirac cone.*

For the time-dependent multilayer under transverse electric polarization (Fig. 8), the 3D dispersion surface $E(k_x, k_y)$ shown in Fig. 8a reveals that temporal modulation of the permittivity introduces noticeable band reshaping compared to the static case. The top-down projection highlights weak anisotropy (Fig. 8b), with slight angular deformation of the band valleys, indicating that the modulation primarily affects the curvature rather than opening a full photonic gap. The angularly resolved dispersions (Fig. 8c), plotted as a function of $\Delta k/(\pi/a)$ for various propagation directions $\phi$, demonstrate that the dynamic modulation induces modest frequency shifts and band splitting near the band edges. The shifts are most pronounced for $\phi = 0°$ and $\phi = 90°$, suggesting that temporal variation couples selectively to longitudinal and transverse components of the in-plane momentum. Despite these modifications, the overall transverse electric band topology remains continuous, confirming that dynamic modulation influences dispersion but does not generate a complete stop band. These results establish that, unlike the transverse magnetic polarization where static dielectric contrast dominates, the transverse electric response remains nearly isotropic and continuous even under time-dependent permittivity variation. A local analysis of the angularly resolved crossings in Fig. 8c shows finite slopes on both intersecting branches and no evidence of a perfectly flat band; thus the observed crossing is consistent with a slightly tilted type-I Dirac cone and does not satisfy the critical tilt condition required for a type-III Dirac point.